\documentclass{article}

\usepackage{PRIMEarxiv}

\usepackage[utf8]{inputenc} 
\usepackage[T1]{fontenc}    
\usepackage[hidelinks]{hyperref}       
\usepackage{url}            
\usepackage{booktabs}       
\usepackage{amsfonts}       
\usepackage{nicefrac}       
\usepackage{microtype}      
\usepackage{fancyhdr}       
\usepackage{graphicx}       
\graphicspath{{media/}}     

\title{Generalised Automatic Anatomy Finder (GAAF):\\A general framework for 3D location-finding in CT scans}

\author{
  Edward G. A. Henderson, Eliana M. Vasquez Osorio, Marcel van Herk, and Andrew F. Green \\
  Division of Cancer Sciences \\
  The University of Manchester \\
  Manchester, UK\\
  \texttt{edward.henderson@postgrad.manchester.ac.uk}
}

\begin{document}
\maketitle

\begin{abstract}
We present GAAF, a \textbf{G}eneralised \textbf{A}utomatic \textbf{A}natomy \textbf{F}inder, for the identification of generic anatomical locations in 3D CT scans. GAAF is an end-to-end pipeline, with dedicated modules for data pre-processing, model training, and inference. At it's core, GAAF uses a custom a localisation convolutional neural network (CNN). The CNN model is small, lightweight and can be adjusted to suit the particular application. The GAAF framework has so far been tested in the head and neck, and is able to find anatomical locations such as the centre-of-mass of the brainstem. GAAF was evaluated in an open-access dataset and is capable of accurate and robust localisation performance.  All our code is open source and available at  \url{https://github.com/rrr-uom-projects/GAAF}.
\end{abstract}

\section{Introduction}
The use of machine learning is growing in popularity in medical imaging. In particular in radiotherapy, convolutional neural networks (CNNs) have been applied to several tasks including outcome prediction, organ-at-risk and target segmentation and image registration. However, full-resolution CT scans are large 3D volumes, which tend to require down-sampling or cropping to a region-of-interest in order for them to be passed through machine learning pipelines, even using modern, high-end hardware.

The loss of image resolution when down-sampling can have undesirable impacts, so cropping to particular regions-of-interest is often applied. However identification of the volume to crop can introduce uncertainty and, when done manually, can considerably slow down a machine learning workflow. Some auto-segmentation methods apply two-stage approaches: first to automatically identify the region-of-interest (by either performing an initial segmentation or bounding-box identification in low resolution and full field-of-view) and then performing the final segmentation at a finer resolution.

We originally developed GAAF for this purpose, as a preprocessing step to identify a single target to act as a crop centre\cite{henderson2021po}. GAAF has now been generalised so a localisation model can be trained for any site and target.

The intended use for GAAF is that the user will bring a dataset of full-resolution CT scans and structure of interest to be located. GAAF has dedicated pre-processing and training modules to prepare the users data and then train a custom localisation model. The GAAF inference module can then be used to locate the target in unseen data. The inference module provides options to either return the raw coordinates or perform a crop around the found location. The latter is useful if GAAF is to be implemented as part of a larger pipeline, e.g. for 3D auto-segmentation of large CT scans.
\newpage
\section{Methodology}
The GAAF project is divided into three modules; data preprocessing, model training and inference. The modules are intended to be used sequentially. 

The preprocessing module is used to prepare the training dataset (presented by the user) for model training. The CT scan will be downsampled to the desired size ($64\times128\times128$ voxels recommended) and the associated target structure (provided in mask form) converted to a heatmap target. The training module uses the preprocessed data to train a custom localisation model. By default, a five-fold cross validation is completed through the training dataset presented. Currently the best of the folds must be manually selected for inference using the included testing functionality.

\subsection{Model details}
\begin{figure}[ht]
\centerline{\includegraphics[width=\textwidth]{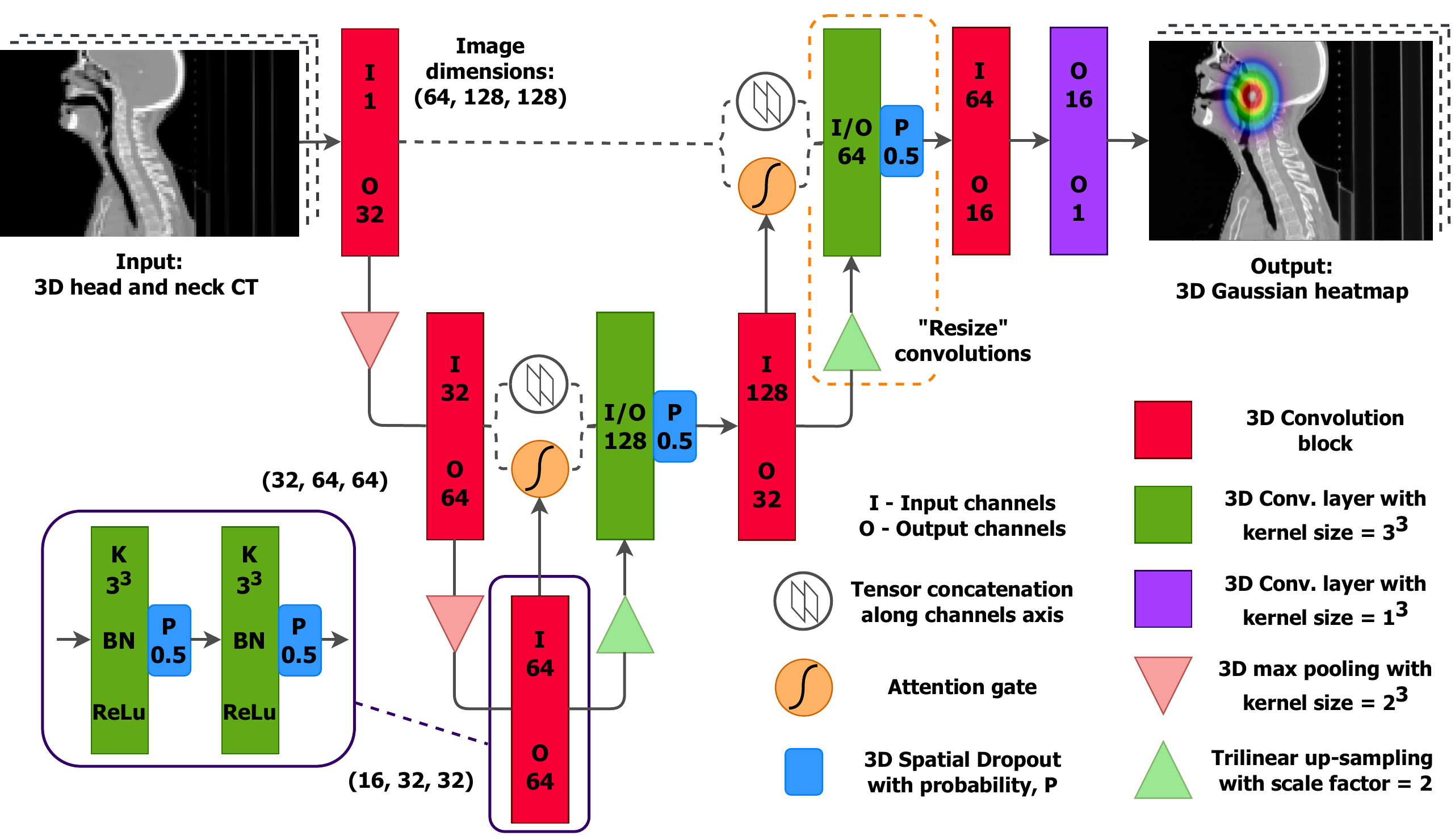}}
\caption{A schematic of the localisation CNN architecture. The model is a compact UNet-like design with an encoder-decoder pathway. The model is trained using heat-map matching to identify a target location within downsampled, but full field-of-view CT images. GAAF includes the option to either use standard UNet-like concatenation skip connections\cite{Ronneberger2015}, or to include attention gates as implemented in \cite{Oktay2018}. In this figure, the example input and output show the model finding the centre-of-mass of the brainstem in a 3D head and neck CT scan.}
\label{full_headHunter_Fig}
\end{figure}

The localisation CNN model is a compact, fully-convolutional UNet-like design and is illustrated in Figure~\ref{full_headHunter_Fig}\cite{Ronneberger2015}. The CNN includes spatial dropout to reduce overfitting, resample convolutions to reduce the impact of checkerboard artefacts in the output and (optionally) atention gates as described by Oktay at al.\cite{Oktay2018}.

The model, trained using supervised learning, produces 3D Gaussian distribution heatmaps centred on the target location. The model is trained to replicate the heatmaps generated in the preprocessing step using a weighted combination of the L2 and L1 loss functions and the Adam optimiser. This heatmap matching technique is similar to the method presented by Payer et al.\cite{Payer2019}. At training time, simple shift and left-right flip augmentations are applied to improve model robustness.

\subsection{Model evaluation}
To evaluate the localisation model, we trained versions to locate the centre-of-mass of the brainstem using a dataset of 185 HN CT images from The Cancer Imaging Archive (TCIA)\cite{HNSCC2020, Elhalawani2017, Grossberg2018, Clark2013}. We performed two separate five-fold cross-validations, one using the localisation architecture with attention gates and one using regular skip connections.

We additionally tested two methods of acquiring the target location from the model output; first by naively taking the maximum value of the heatmap, and second by fitting a 3D Gaussian distribution to the output and taking the expectation value. Each of the estimated set of location coordinates were scaled back into the frame-of-reference of full resolution CT image. We then measured the 3D Euclidean distance between the predicted location and the true location of the centre-of-mass of the brainstem.

\section{Evaluation results}
\begin{figure}[ht]
\centerline{\includegraphics[width=\textwidth]{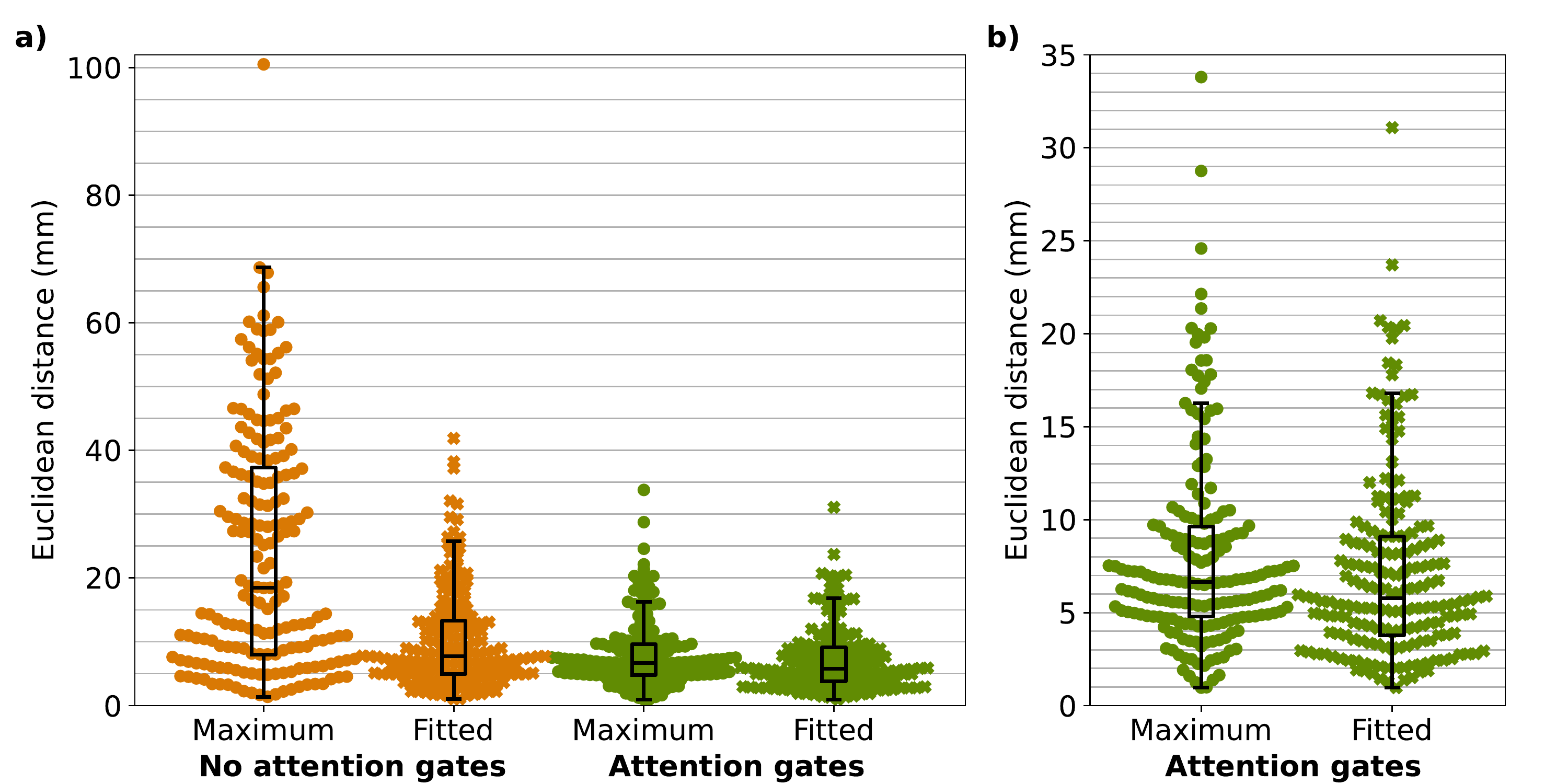}}
\caption{a) Results of the five-fold cross-validation for the localisation models with (green) and without attention gates (orange). We show swarm-plots, with box and whiskers, of the Euclidean distance to the gold-standard target for when the location is estimated by taking the maximum value of the output heatmap (circles) and when estimated by taking the expectation value of a fitted 3D Gaussian distribution (crosses). b) A zoomed-in view of the same results for the localisation model with attention gates.}
\label{headHunter_results_Fig}
\end{figure}

The five-fold cross-validation results for the localisation model with attention gates are shown in Figure~\ref{headHunter_results_Fig}a. The median 3D Euclidean distance to the gold-standard target was $6.7\pm5.3$ mm when taking the heatmaps maximum value and $5.8\pm5.0$ mm when taking the expectation of a fitted Gaussian distribution (equivalent to $\sim$2-3 voxels).

The corresponding five-fold cross-validation results for the localisation model without attention gates are shown in Figure~\ref{headHunter_results_Fig}b. The median 3D Euclidean distance to the gold-standard target was $18.5\pm18.8$ mm when taking the maximum and $7.7\pm7.8$ mm when taking the expectation of a fitted distribution.

The location-finding results for the models containing attention gates were far superior to those without attention gates. We therefore advise GAAF users to use the architecture variant with attention gates. 

The difference in the results between the methods for acquiring the predicted location from the output target heatmap are very small when using attention gates. However, fitting a 3D Gaussian distribution to the output and taking the expectation value appears to be slightly advantageous and is anticipated to be more robust in practice. As a result, the GAAF inference module estimates the predicted location coordinates in this way.

\section{Conclusion}
We have presented GAAF, a generalised location finding network. This package is multi-purpose and generalised to be adapted to CT scans of any anatomical site. GAAF can be used individually to identify locations, or as a part of a wider machine learning pipeline, to crop anatomically consistent sub-volumes from full field-of-view CT scans.

\bibliographystyle{ama}  
\bibliography{references}  

\end{document}